\documentclass[aps,prl,twocolumn,superscriptaddress]{revtex4}
\usepackage{amsfonts}
\usepackage{amsmath}
\usepackage{graphicx}
\usepackage{dsfont}
\usepackage{diagbox}
\usepackage{array}
\usepackage{amssymb}
\usepackage{bbm}
\usepackage{float}
\usepackage{subfigure}
\usepackage{url}
\usepackage{hyperref}

\newcommand{\PreserveBackslash}[1]{\let\temp=\\#1\let\\=\temp}
\newcolumntype{C}[1]{>{\PreserveBackslash\centering}p{#1}}
\newcolumntype{R}[1]{>{\PreserveBackslash\raggedleft}p{#1}}
\newcolumntype{L}[1]{>{\PreserveBackslash\raggedright}p{#1}}

\begin{document}

\title{Self-doped Mott insulator for parent compounds of nickelate
superconductors}
\author{Guang-Ming Zhang}
\email[]{gmzhang@tsinghua.edu.cn}
\affiliation{State Key Laboratory of Low-Dimensional Quantum Physics and Department of
Physics, Tsinghua University, Beijing 100084, China}
\affiliation{Frontier Science Center for Quantum Information, Beijing 100084, China}
\author{Yi-feng Yang}
\email[]{yifeng@iphy.ac.cn}
\affiliation{Beijing National Lab for Condensed Matter Physics and Institute of Physics,
Chinese Academy of Sciences, Beijing 100190, China}
\affiliation{School of Physical Sciences, University of Chinese Academy of Sciences,
Beijing 100190, China}
\affiliation{Songshan Lake Materials Laboratory, Dongguan, Guangdong 523808, China}
\author{Fu-Chun Zhang}
\email[]{fuchun@ucas.ac.cn}
\affiliation{Kavli Institute for Theoretical Sciences and CAS Center for Topological
Quantum Computation, University of Chinese Academy of Sciences, Beijing
100190, China }
\date{\today }

\begin{abstract}
We propose the parent compound of the newly discovered superconducting
nickelate Nd$_{1-x}$Sr$_{x}$NiO$_{2}$ as a self-doped Mott insulator, in
which the low-density Nd-$5d$ conduction electrons couple to localized Ni-3$%
d_{x^{2}-y^{2}}$ electrons to form Kondo spin singlets at low temperatures.
This proposal is motivated with our analyses of the reported resistivity and
Hall coefficient data in the normal state, showing logarithmic temperature
dependence at low temperatures. In the strong Kondo coupling limit, we
derive a generalized $t$-$J$ model with both Kondo singlets and nickel
holons moving through the lattice of otherwise nickel spin-1/2 background.
The antiferromagnetic long-range order is therefore suppressed as observed
in experiments. With Sr-doping, the number of holons on the nickel sites
increases, giving rise to the superconductivity and a strange metal phase
analogous to those in superconducting copper oxides.
\end{abstract}

\maketitle

\textbf{Introduction.} - Recent discovery of superconductivity in Nd$_{0.8}$%
Sr$_{0.2}$NiO$_{2}$ \cite{Li2019} has stimulated intensive interest in
understanding its pairing mechanism, in particular, its similarity and
difference compared to that in cuprate superconductors \cite%
{Norman2019,Sawatzky2019,Kuroki2019,Hepting2019,Normura2019,Hongming2019}.
Despite tremendous efforts over past thirty years, high $T_{c}$
superconductivity (SC) remains one of the most challenging topics in
condensed matter physics \cite{Muller,Anderson,AtoZ,LeeNagaosaWen}. The
parent compounds of copper oxides may be described as a Mott insulator with
antiferromagnetic (AF) long-range order. Superconductivity arises when
additional holes are introduced on the oxygen sites in the CuO$_{2}$ planes
upon chemical doping. These holes combine with the $3d_{x^{2}-y^{2}}$ spins
of Cu-ions to form the Zhang-Rice singlets moving through the square lattice
of Cu-ions by the exchange with their neighboring Cu-spins, which leads to
an effective two-dimensional $t-J$ model to describe the low-energy physics
of the cuprates \cite{ZhangRice}. The AF order is destroyed rapidly by small
hole doping, while at optimal doping, the $d$-wave SC is established in bulk
cuprates \cite{Shen,Harlingen,Tsuei}. It has been a long-standing question
if these ``cuprate-Mott" conditions can be realized in other oxides.
Extensive efforts have been made to investigate the nickel oxides both
theoretically and experimentally \cite%
{Anisimov1999,Hayward1999,Lee2004,Botana2017,Chaloupka2008,Hansmann2009,Middey2016,Boris2011,Benckiser2011,Disa2015,Zhang2017}%
.

Single crystal thin films of infinite-layer nickelates were lately
synthesized using soft-chemistry topotactic reduction. Superconductivity was
reported below $9\sim 15\,$K in the hole-doped Nd$_{0.8}$Sr$_{0.2}$NiO$_{2}$
\cite{Li2019}. The nickelate superconductors have similar crystal structure
as cuprates, and the monovalent Ni$^{1+}$-ions also possess the same $3d^{9}$
configuration as Cu$^{2+}$-ions. It is therefore thought to be the same as
cuprates. However, the parent compound NdNiO$_{2}$ displays metallic
behavior at high temperatures with a resistivity upturn below about $70$ K,
and shows no sign of any magnetic long-range order in the whole measured
temperature range \cite{Hayward2003}. Similar results have also been found
previously in LaNiO$_{2}$ \cite{Ikeda2016}. These experimental observations
are in sharp contrast with the naive expectation of a Mott insulator with AF
long-range order for the parent compounds of nickelates. It is therefore
important to address what is the nature of the parent compounds and how the
AF long-range order is suppressed.

\begin{figure}[t]
\centerline{{\includegraphics[width=.45\textwidth]{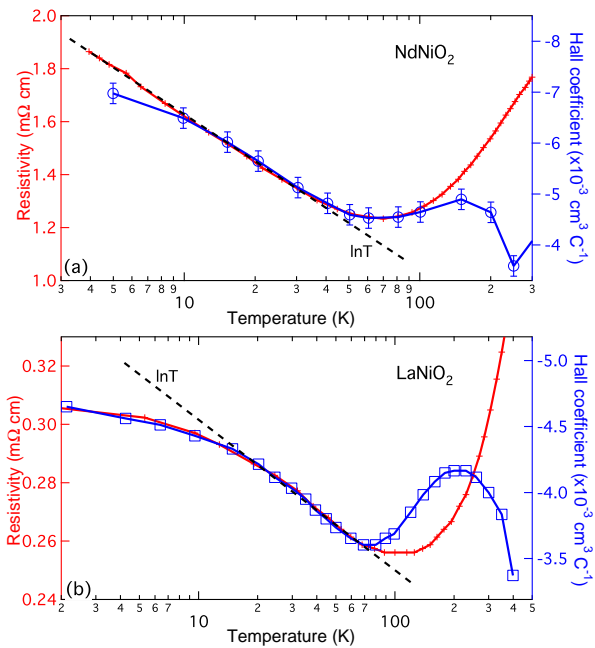}}}
\caption{Logarithmic temperature dependence of the resistivity (red color)
and the Hall coefficient (blue color) at low temperatures for (a) NdNiO$_{2}$
with the experimental data adopted from Ref.~\protect\cite{Li2019}; (b) LaNiO%
$_{2}$ reproduced from Ref.~\protect\cite{Ikeda2016}. The dashed lines are
the ${\rm ln} T$ fits.}
\label{fig1}
\end{figure}

\textbf{Key experimental evidences.} - Figure~\ref{fig1} presents the
resistivity and Hall data as functions of temperature for both parent
compounds NdNiO$_{2}$ and LaNiO$_{2}$. Surprisingly, when the data were put
on a linear-log scale, we find that the resistivity $\rho $ upturn well
obeys a logarithmic temperature ($\ln T$) dependence below about $40$ K down
to $4$ K for NdNiO$_{2}$ and below about $70$ K down to $11$ K for LaNiO$%
_{2} $. This is a clear evidence of magnetic Kondo scattering \cite%
{HewsonBook,Foot-Note-2}.

This Kondo scenario is further supported by the Hall effect data in the both
compounds. While the Hall coefficient $R_{H}$ exhibits non-monotonic
temperature dependence, very different from that of the resistivity in the
high temperature metallic regime, it shows the same $\ln T$ dependence at
low temperatures. In the Kondo systems, we have $R_{H}\propto \rho $, due to
the incoherent skew scattering associated with the localized magnetic
impurity \cite{Fert1987,Nagaosa2010}. Thus both the resistivity and Hall
coefficient support the presence of the magnetic Kondo scattering in the
parent compounds of nickelate superconductors. Moreover, at high
temperatures where the skew scattering is negligible and the normal Hall
effect dominates, the magnitude of the Hall coefficient is found to be only
about $-4\times 10^{-3}\,$cm$^{3}$\thinspace C$^{-1}$ for NdNiO$_{2}$ and $%
-3\times 10^{-3}\,$cm$^{3}$\thinspace C$^{-1}$ for LaNiO$_{2}$. Both are an
order of magnitude higher than those of normal heavy fermion metals. For
example, we have $R_{H}\approx -3.5\times 10^{-4}\,$cm$^{3}$\thinspace C$%
^{-1}$ in all three Ce$M$In$_{5}$ compounds ($M$= Co, Rh, Ir) at high
temperatures \cite{Hundley2004}. This indicates that there are only a few
percent of electron-like carriers per unit cell in both NdNiO$_{2}$ and LaNiO%
$_{2}$. Therefore, the parent compounds of nickelates belong to a Kondo
system with low-density charge carriers.

Below we examine the Kondo scenario for NdNiO$_{2}$ from the microscopic
picture. The first-principles band structure calculations \cite%
{Lee-Pickett2004} show that the Nd $5d$ orbitals in NdNiO$_{2}$ are
hybridized with the Ni $3d$ orbitals, leading to small Fermi pockets of
dominantly Nd $5d$ electrons in the Brillouin zone. Nd-5d conduction
electrons have a low electron density of $n_{c}\ll 1$ per Ni-site, coupling
to the localized Ni$^{1+}$ spin-1/2 of $3d_{x^{2}-y^{2}}$ orbital to form
Kondo spin singlets (doublons) \cite{Foot-Note-1}. Here we have considered
Ni-$3d_{x^{2}-y^{2}}$ electrons to be strongly correlated with a large
on-site Coulomb repulsion $U$ to disfavor double occupation on the same
sites.

With this picture in mind, it is attempted to propose a Kondo Hamiltonian to
describe the parent compounds of nickelates. However, unlike the usual Kondo
lattice model, the Ni$^{1+}$ localized spins here are coupled mainly by
superexchange interaction through the O-$2p$ orbitals as same as in the
cuprates, though the coupling on nickel sites is small. Thus the starting
point should actually be a background lattice of Ni$^{1+}$ localized spins
with the nearest-neighbor AF Heisenberg superexchange coupling and
additional local Kondo exchanges with the itinerant $5d$ electrons.

For the parent compound, we have correspondingly $1-n_{c}$ electrons per
Ni-site, or $n_{c}N_{s}$ ($N_{s}$ as the total number of Ni-sites) empty
nickel sites (holons) on the NiO$_{2}$ plane. This introduces a strongly
renormalized hopping term of holons. A schematic picture is displayed in
Fig.2 The presence of both the Kondo singlets/doublons and the holons can
suppress very efficiently the AF long-range order and cause a phase
transition from the Mott insulating state to a metallic state. Actually, as
we will show below, an effective low-energy model Hamiltonian can be derived
in terms of the doublons, holons and localized spins, describing a
self-doped Mott metallic state even in the parent LnNiO$_{2}$ (Ln=La, Nd)
compounds. Upon further Sr hole doping, such a low-energy effective model is
expected to exhibit $d$-wave pairing instability as in the usual $t$-$J$
model.

\begin{figure}[t]
\centerline{{\includegraphics[width=.45\textwidth]{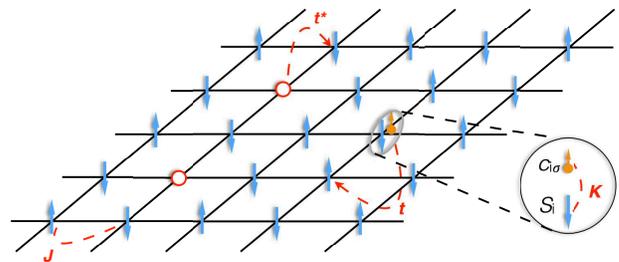}}}
\caption{Illustration of the effective model on a two-dimensional square
lattice of NiO$_{2}$ plane of NdNiO$_{2}$. Blue arrow represents Ni-spin,
which interacts with its neighboring spin antiferromagetically by coupling $J
$. Orange arrow denotes Nd-$5d$ electron, which couples to Ni-spin by the
Kondo coupling $K$, to form a Kondo singlet (doublon). Red circle represents
Ni-$3d^{8}$ configuration, or a holon. $t$ and $t^{\ast }$ are the hoping
integrals of doublon and holon, respectively. Not shown is the holon-doublon
anhilation into a Ni-spin.}
\label{fig2}
\end{figure}

\textbf{Effective model Hamiltonian.} - We consider Ni-$3d^{8}$ and Nd-$%
5d^{0}$ as the vacuum, and start with the localized $3d_{x^{2}-y^{2}}$ spins
on the NiO$_{2}$ plane that form a two-dimensional quantum Heisenberg model
with nearest neighbour AF superexchange interactions,
\begin{equation}
H_{J}=J\sum_{\langle ij\rangle }S_{i}\cdot S_{j}.
\end{equation}%
This is similar to the cuprates, where the superexchange interaction is
induced by the O-$2p$ orbitals and the parent compound is a Mott insulating
state with AF long-range orders. In nickelates, however, we have to further
consider the Kondo coupling with the Nd or La $5d$ conduction electrons.
This leads to the following Kondo lattice Hamiltonian,
\begin{equation}
H_{K}=-t\sum_{\langle ij\rangle ,\sigma }\left( c_{i\sigma }^{\dagger
}c_{j\sigma }+h.c.\right) +\frac{K}{2}\sum_{j\alpha ;\sigma \sigma ^{\prime
}}S_{j}^{\alpha }c_{j\sigma }^{\dagger }\tau _{\sigma \sigma ^{\prime
}}^{\alpha }c_{j\sigma ^{\prime }},
\end{equation}%
where $t$ describes the effective hoping amplitude of the $5d$ itinerant
electrons projected on the square lattice sites of the Ni$^{1+}$ ions, and $%
\tau ^{\alpha }$ ($\alpha =x,y,z$) are the spin-1/2 Pauli matrices. We
consider a single $5d$ orbital for Nd for simplicity. For a low-carrier
density system, the average number of conduction electrons is very small,
i.e., $N_{s}^{-1}\sum_{j\sigma }\langle c_{j\sigma }^{\dagger }c_{j\sigma
}\rangle =n_{c}\ll 1$.

In the parent compound LnNiO$_{2}$ (Ln=La, Nd), the total electron density
is $1$ per unit cell, hence the total holon density $n_{h}=n_{c}$. For Sr
doped compounds, we have $\delta =n_{h}-n_{c}>0$. To describe the doping
effect, we introduce the pseudofermion representation for the spin-1/2 local
moments,
\begin{equation*}
S_{j}^{+}=f_{j\uparrow }^{\dagger }f_{j\downarrow },S_{j}^{-}=f_{j\downarrow
}^{\dagger }f_{j\uparrow },S_{j}^{z}=\frac{1}{2}\left( f_{j\uparrow
}^{\dagger }f_{j\uparrow }-f_{j\downarrow }^{\dagger }f_{j\downarrow
}\right) ,
\end{equation*}%
where $f_{j\sigma }$ is a fermionic operator and denotes a spinon on site $j$%
. The holon hopping term between empty nickel sites is then given by
\begin{equation}
H_{t^{\ast }}=-t^{\ast }\sum_{\langle ij\rangle ,\sigma }\left(
h_{i}f_{i\sigma }^{\dagger }f_{j\sigma }h_{j}^{\dagger }+h.c.\right) ,
\end{equation}%
where $h_{j}^{\dagger }$ is the bosonic operator creating a holon on the $j$%
-th site. In this representation, the Ni $3d_{x^{2}-y^{2}}$ electron
operator is given by $d_{j\sigma }=h_{j}^{\dagger }f_{j\sigma }$ with a
local constraint, $h_{j}^{\dagger }h_{j}+\sum_{\sigma }f_{j\sigma }^{\dagger
}f_{j\sigma }=1$. This is just the slave-boson representation for the
constrained electrons without double occupancy.

All together, the total model Hamiltonian for nickelates consists of three
terms,
\begin{equation}
H=H_{J}+H_{K}+H_{t^{\ast }}.
\end{equation}%
This model contains several key energy scales. While the electron hopping $t$
may be roughly estimated from band calculations, the holon hopping $t^{\ast }
$ is strongly renormalized due to the background AF correlations and thus
contribute little to the transport measurements in the parent compounds. The
kinetic energy in the Hamiltonian is therefore relatively small due to the
small number of charge carriers without Sr doping. The Heisenberg
superexchange $J$ is also expected to be smaller (possibly the order of $10$
meV) compared to that (about $100$ meV) in cuprates due to the larger charge
transfer energy between O-$2p$ and Ni-$3d_{x^{2}-y^{2}}$ orbitals. Actually
the Heisenberg exchange energy is further reduced in a paramagnetic
background. For the Kondo temperature of the value of $10$ K or $1$ meV,
which is about one tenth of the temperature of resistivity minimum in both
LaNiO$_{2}$ and NdNiO$_{2}$, a Kondo coupling of roughly the order of $100$ $%
{\text{m}}${eV} would be expected for a low-carrier density system with a
small electron density of states \cite{Yifeng}. Thus for the parent
compounds of nickelates, the Kondo coupling is a relatively large energy
scale in the above model Hamiltonian.

From these analyses, one may anticipate that the ground state of the
nickelate parent compounds may be to some extent captured by the large $K$
limit of the Hamiltonian. The Kondo singlets are then well established
between the Ni $3d_{x^{2}-y^{2}}$ spins and the $5d$ conduction electrons.
To explore this possibility, we introduce the doublon operators for the
on-site Kondo spin singlet and triplets:
\begin{eqnarray*}
b_{j0}^{\dagger } &=&\frac{1}{\sqrt{2}}\left( f_{j\uparrow }^{\dagger
}c_{j\downarrow }^{\dagger }-f_{j\downarrow }^{\dagger }c_{j\uparrow
}^{\dagger }\right) ;\text{ } \\
b_{j1}^{\dagger } &=&f_{j\uparrow }^{\dagger }c_{j\uparrow }^{\dagger },\
b_{j2}^{\dagger }=\frac{1}{\sqrt{2}}\left( f_{j\uparrow }^{\dagger
}c_{j\downarrow }^{\dagger }+f_{j\downarrow }^{\dagger }c_{j\uparrow
}^{\dagger }\right) ,\ b_{j3}^{\dagger }=f_{j\downarrow }^{\dagger
}c_{j\downarrow }^{\dagger }.
\end{eqnarray*}%
The Kondo exchange term is then transformed to,
\begin{equation}
\frac{K}{2}\sum_{j\alpha ;\sigma \sigma ^{\prime }}S_{j}^{\alpha }c_{j\sigma
}^{\dagger }\tau _{\sigma \sigma ^{\prime }}^{\alpha }c_{j\sigma ^{\prime }}=%
\frac{K}{4}\sum_{\mu =1}^{3}b_{j\mu }^{\dagger }b_{j\mu }-\frac{3K}{4}%
\sum_{j}b_{j0}^{\dagger }b_{j0},
\end{equation}%
which describes the doublon formation on each site, namely, the Kondo
singlet or triplet pair formed by each conduction electron with the
localized spinon. However, the Kondo triplet costs a larger energy of $K$
and is therefore not favored. In addition, there can also be three-electron
states with two conduction electrons and the localized spinon on the same
site, $e_{j\sigma }^{\dagger }=f_{j\sigma }^{\dagger }c_{j\uparrow
}^{\dagger }c_{j\downarrow }^{\dagger }$, and one-electron states with the
unpaired spinon only, $\widetilde{f}_{j\sigma }=\left( 1-n_{j}^{c}\right)
f_{j\sigma }$. So these operators should be used with the constraint
\begin{equation}
h_{j}^{\dagger }h_{j}+\sum_{\mu =0}^{3}b_{j\mu }^{\dagger }b_{j\mu
}+\sum_{\sigma }\left( \widetilde{f}_{j\sigma }^{\dagger }\widetilde{f}%
_{j\sigma }+e_{j\sigma }^{\dagger }e_{j\sigma }\right) =1,
\end{equation}%
for each site. These new operators do not commute in a simple way so for
simplicity we should avoid direct operation using their commutation
relations.

In the large $K$ limit, following the method used in Ref. \cite%
{Lacroix1985,Sigrist1992}, a low-energy effective Hamiltonian can be derived
by first rewriting the hopping term $H_{t}$ in terms of the new operators $%
b_{j0}$, $b_{j\mu }$ ($\mu =1,2,3$), $e_{j\sigma }$, and $\widetilde{f}%
_{j\sigma }$ and then employing the canonical transformation, $H_{\mathrm{eff%
}}=e^{-S}He^{S}$, to eliminate all high-energy terms containing $b_{j\mu }$ (%
$\mu =1,2,3$) and $e_{j\sigma }$ while keeping only the on-site doublon ($%
b_{j0}$) and unpaired spinons ($\widetilde{f}_{j\sigma }$). In the infinite-$%
K$ limit, in particular, the low-energy effective model becomes a simple
form
\begin{eqnarray}
H_{\mathrm{eff}} &=&-t^{\ast }\sum_{\langle ij\rangle ,\sigma }\left( h_{i}%
\widetilde{f}_{i\sigma }^{\dagger }\widetilde{f}_{j\sigma }h_{j}^{\dagger
}+h.c.\right) +J\sum_{\langle ij\rangle }\widetilde{S}_{i}\cdot \widetilde{S}%
_{j}  \notag \\
&&-\frac{t}{2}\sum_{\langle ij\rangle ,\sigma }\left( b_{i0}^{\dagger }%
\widetilde{f}_{i\sigma }\widetilde{f}_{j\sigma }^{\dagger
}b_{j0}+h.c.\right) ,
\end{eqnarray}%
where the spin operators are expressed as $\widetilde{S}_{j}^{\alpha
}=\sum_{\sigma \sigma ^{\prime }}\widetilde{f}_{j\sigma }^{\dagger }\frac{1}{%
2}\tau _{\sigma \sigma ^{\prime }}^{\alpha }\widetilde{f}_{j\sigma ^{\prime
}}$ with a local constraint $h_{j}^{\dagger }h_{j}+b_{j0}^{\dagger
}b_{j0}+\sum_{\sigma }\widetilde{f}_{j\sigma }^{\dagger }\widetilde{f}%
_{j\sigma }=1$. For a large but finite $K$, apart from some complicated
interactions, an additional term should be included
\begin{equation}
H_{b}=-\frac{3}{4}\left( K+\frac{t^{2}}{K}\right) \sum_{j}b_{j0}^{\dagger
}b_{j0}+\frac{5t^{2}}{12K}\sum_{\langle ij\rangle }b_{i0}^{\dagger
}b_{i0}b_{j0}^{\dagger }b_{j0},
\end{equation}%
which could be used to describe the doublon condensation.

\textbf{Discussions} - The above effective low-energy Hamiltonian is very
similar to the usual $t$-$J$ model for cuprates \cite{ZhangRice}, except
that it includes two different types of charge carriers: the Kondo singlets
(doublons) and the holons on the Ni sites. Their presence can efficiently
suppress the AF long-range order and bring the phase transition from a Mott
insulator to a self-doped Mott metallic state. The effective model therefore
describes a self-doped Mott insulating state as the parent state of
nickelate superconductors, with possibly an enhanced effective mass for the
charge carriers. It also provides an interesting example of holon-doublon
excitations for destroying the Mott insulator, although the doublons here
are associated with the Kondo singlets rather than doubly occupied Ni $%
3d_{x^{2}-y^{2}}$ orbitals. At high temperatures, the doublons become
deconfined, causing incoherent Kondo scattering as observed in experiments.

Furthermore, the Sr hole doping reduces the number of electron carriers and
thus suppresses the contribution of doublons. At large doping, the effective
model is then reduced to the usual $t$-$J$ model. In cuprates, the Cu $%
3d_{x^{2}-y^{2}}$ orbitals and the O-$2p$ orbitals are strongly hybridized.
The doped holes sit on the oxygen sites, forming the Zhang-Rice singlets
with Cu$^{2+}$ localized spins. By contrast, the holes in nickelates reside
on the Ni-ions, leading to a spin zero state or holon due to the much less
overlap with the O-$2p$ band \cite{Sawatzky2019}. Sr doping hence introduces
extra holes on the Ni sites, which further drives the system away from the
AF Mott insulating phase, resembling that in the optimal or overdoped
cuprates. However, even at $20$\% Sr doping, the electron carriers are still
present, as manifested by the negative Hall coefficient at high temperatures
in Nd$_{0.8}$Sr$_{0.2}$NiO$_{2}$ \cite{Li2019}. Since the electron carrier
density is reduced with hole doping, the smaller magnitude of $R_{H}$ in Nd$%
_{0.8}$Sr$_{0.2}$NiO$_{2}$ cannot be explained by a single carrier model but
rather indicates a cancellation of electron and hole contributions. The
latter grows gradually with Sr doping and eventually becomes dominant at low
temperatures in Nd$_{0.8}$Sr$_{0.2}$NiO$_{2}$, causing the sign change of
the Hall coefficient below about $50$ K.

Experimentally, with $20$\% Sr doping in NdNiO$_{2}$, superconductivity also
emerges and has the highest transition temperature of about $15$ K.
Interestingly, when fitted with a power-law temperature dependence, $\rho
\propto T^{\alpha }$, we notice for this particular sample that the electric
resistivity exhibits a non-Fermi liquid behavior in the normal state.
Actually, an excellent agreement could be obtained with $\alpha =1.13\pm 0.02
$ over a wide range from slightly above the superconducting transition
temperature up to the room temperature. In fact, for all reported samples
with high superconducting transition temperature, a good power-law fit can
always be obtained with $\alpha \approx 1.1\sim 1.3$. This is reminiscent to
the optimal doped cuprate superconductors and suggests a similar strange
metal phase for the normal state of optimal doped nickelate superconductors.

\textbf{Conclusion and outlook} - Our proposed model bridges the Kondo
lattice model for heavy fermions and the $t$-$J$ model for cuprates.
However, it is different from both models in the sense that it combines some
new physics that is not included in either of them. Unlike the usual Kondo
lattice system, the exchange interaction here between localized spins are
produced by the superexchange coupling rather than the RKKY coupling. Thus
at low carrier density, the magnetic ground state is not ferromagnetic as
one would expect for the Kondo lattice. On the other hand, the nickelate
system indeed exhibits incoherent Kondo scatterings as revealed in the
transport properties at high temperatures. Unlike cuprates, the presence of
strong Kondo coupling could lead to holon-doublon excitations even in the
parent compound. This self-doping effect suppresses the AF long-range order
and produces the paramagnetic metallic ground state. The parent compound of
nickelate superconductor is therefore described as a self-doped Mott state.
This makes it somehow different from the cuprates but resembles certain
organic superconductors under pressure, which reduces the on-site Coulomb
repulsion $U$ and induces a transition from Mott insulator to gossamer
superconductor with both holons and doublons \cite{gossamer}.

\textit{Acknowledgment}.- The authors would like to acknowledge the
discussions with Wei-Qiang Chen, Hong-Ming Weng, and Yu Li. This work was
supported by the National Key Research and Development Program of MOST of
China (2016YFYA0300300, 2017YFA0302902, 2017YFA0303103), the National
Natural Science Foundation of China (11774401 and 11674278), the State Key
Development Program for Basic Research of China (2014CB921203 and 2015CB921303), 
and the Strategic Priority Research Program of CAS (Grand No. XDB28000000).


\begin{thebibliography}{99}
\bibitem{Foot-Note-2} Actually an alternative explanation for the
resistivity upturn at low temperatures is weak localization due to the
presence of disorder holes in the NiO$_{2}$ plane. It can also give rise to
a logarithmic temperature dependent correction. However, the corresponding
correction to the Hall coefficient is independent of temperature in the same
region, which does not support this explanation.
\bibitem{Foot-Note-1} The presence of magnetic impurities may be at first
glance ascribed to the Nd $4f$ moments. However, the Nd$^{3+}$ ion contains
three $f$ electrons forming a localized spin-3/2 moment, which acts more
like a classical spin as in manganites and therefore disfavors spin-flip
scattering as the quantum spin-1/2 moment. Their energy level is also far
away from the Fermi energy, so it is reasonable to ignore the Nd $4f$
electrons.
\end{thebibliography}
\end{document}